\documentclass[11pt,superscriptaddress,aps,prd,preprint]{revtex4}

\everymath{\displaystyle}
\usepackage{graphicx}

\usepackage[T1]{fontenc}
\usepackage{amsmath}
\usepackage{amssymb}
\usepackage{graphicx}
\usepackage{slashed}

\newcommand{\bea}{\begin{eqnarray}}
\newcommand{\eea}{\end{eqnarray}}

\begin{document}

\title{Casimir effect in Very Special Relativity at finite temperature}

\author{A. F. Santos}
\email{alesandroferreira@fisica.ufmt.br}
\affiliation{Instituto de F\'{\i}sica, Universidade Federal de Mato Grosso,\\
78060-900, Cuiab\'{a}, Mato Grosso, Brazil}

\author{Faqir C. Khanna \footnote{Professor Emeritus - Physics Department, Theoretical Physics Institute, University of Alberta\\
Edmonton, Alberta, Canada}}
\email{khannaf@uvic.ca; fkhanna@ualberta.ca}
\affiliation{Department of Physics and Astronomy, University of
Victoria,BC V8P 5C2, Canada}
%\affiliation{TRIUMF, 4004, Westbrook Mall, Vancouver, British Columbia V6T 2A3, Canada}

%%%%%%%%%%%%%%%%%%%%%%%%%%%%%%%%%%%%%%%%%%%%%%%%%%%%%%%%%%%%%%%%%%%%%%%%%%%%%%%%%%%%%
\begin{abstract}

Recently a great deal of interest in field theories formulated in a Lorentz violating framework has been developed. Here the Very Special Relativity (VSR) is considered. The main aspect in the VSR proposal is that laws of physics are invariant under the subgroups of the Poincaré group preserving the basic elements of special relativity. An important point is that the photon acquires a mass. In this context, the energy-momentum tensor for the electromagnetic field is calculated. From this, the Stefan-Boltzmann law and Casimir effect at finite temperature in VSR are obtained. The effects of temperature are introduced using the Thermo Field Dynamics (TFD) formalism. A comparative analysis with the Casimir effect for the standard electromagnetic case is developed.

\end{abstract}

\maketitle

\section{Introduction}

The model that describes all elementary particles and their interactions, except gravity, is known as the Standard Model (SM). In the last decades, great progress on the theoretical and experimental understanding of the SM have been achieved. This led to highly accurate data that established tight bounds in search for a physics beyond the SM. The main symmetry of the SM is Lorentz symmetry. Although the  violation of Lorentz symmetry has not been observed, many theories dealing with physics beyond the SM predict breaking of Lorentz invariance at Planck scale energy ($\sim 10^{19} \mathrm{GeV}$). The subject of Lorentz violation has been motivated by different arguments in a broad spectrum of theories. Some examples of models that present a Lorentz symmetry break are, string theory \cite{Samuel}, loop quantum gravity \cite{LQG}, non-commutative field theory \cite{NC}, standard model extension, which contains the SM, general relativity and all possible operators that break Lorentz symmetry \cite{SME1, SME2}, a Chern-Simons term in $3+1$ dimension \cite{Carroll} and Very Special Relativity (VSR) \cite{Cohen1, Cohen2}. Here the VSR is considered.

The VSR is based on the idea that Lorentz symmetry is not a fundamental symmetry of nature, but this function is reserved for one of its proper subgroups. In this context, the basic elements of special relativity like time dilation, relativistic velocity addition and a maximal isotropic speed of light are preserved. In addition, modified gauge symmetry is present which admits a variety of new gauge invariant interactions. There are two subgroups fulfilling these requirements: Similitude group ($SIM(2)$) and Homothety group ($HOM(2)$).  The $HOM(2)$ has three parameters and is generated by $T_1=K_x+J_y$, $T_2=K_y-J_x$ and $K_z$, wiht $\vec{J}$ and $\vec{K}$ being the generators of rotations and boosts, respectively. The $SIM(2)$ consists of $HOM(2)$ group plus the $J_z$ generator. It is important to note that, these subgroups preserve the direction of a light-like four-vector $n_\mu$. Then a preferred direction in Minkowski space-time is exhibited in theories that are invariant under these subgroups. In this framework, non-local terms that violate Lorentz symmetry can be constructed as ratios of contractions of $n_\mu$ with other kinematic vectors. These non-local terms that violate Lorentz symmetry are invariant under $HOM(2)$ or $SIM(2)$. Furthermore, all local operators preserving $HOM(2)$ or $SIM(2)$ also preserve Lorentz symmetry. 

Many interesting theoretical and phenomenological aspects of VSR effects have been explored. For example: (i) generation of a neutrino mass without lepton number violation or sterile neutrinos \cite{Cohen2}; (ii) analysis of quantum field theory description for fermions and bosons \cite{Lee}; (iii) VSR extension of supersymmetry \cite{super, super1}; (iv) VSR subgroup in the Moyal noncommutative space-time \cite{Jab}; (v) VSR correction to the Thomas precession \cite{Riv}; (vi) non-abelian gauge transformation \cite{Riv1}; (vii) modification of Maxwell-Chern-Simons electrodynamics \cite{Bufalo}; (viii) thermodynamical properties of the quantum electrodynamics \cite{Bufalo2}; (ix) differential cross section for Bhabha and Compton scattering for the quantum electrodynamics defined with SIM(2) \cite{Bufalo3}; (x) three-body decay process \cite{Tri}; (xi) description of the axion electrodynamics \cite{axions}, among others. However no attention has been made to analyse and study the Casimir effect with the VSR field theories. The present work would deal with Casimir effect at zero and finite temperature.

The Casimir effect \cite{Casimir} is the best known physical manifestation of the quantum vacuum fluctuations. For the electromagnetic field, it consists in the attraction between two parallel metallic plates embedded into the vacuum. The attraction between plates is the result of electromagnetic modes due to boundary conditions or topological effects. This effect was confirmed experimentally with a high degree of accuracy \cite{Sparnaay, Lamoreaux, Mohideen}. Although the Casimir effect was initially studied for the electromagnetic field, it is known to occur for any quantum field under certain boundary condition or topological effects. Here the Casimir effect is calculated for the electromagnetic field in the context of the VSR at finite temperature. The effects of temperature is introduced using the Thermo Field Dynamics (TFD) formalism.

The effects of temperature on a quantum field theory can be introduced through two formalism, imaginary-time formalism or real-time formalism. The Matsubara formalism \cite{Matsubara} is based on a formal substitution of time $t$ by a complex time, i.e., $i\tau$. In this framework, the temperature emerges as a consequence of a compactification of the field in a finite interval on the time axis, $0<\tau<\beta$, where $\beta$ is related to the inverse of temperature. On the other hand, TFD is a real-time finite temperature formalism \cite{Umezawa1, Umezawa2, Umezawa22, Khanna1, Khanna2}. In this formalism the thermal average of any operator {\cal O} is equal to its temperature dependent vacuum expectation value, i.e., $\langle O \rangle=\langle 0(\beta)| O|0(\beta) \rangle$, where $|0(\beta) \rangle$ is a thermal vacuum, with $\beta\propto\frac{1}{T}$, and $T$ begin the temperature. In order to satisfy this relationship, two elements are needed. The first one is a doubling in the Hilbert space giving rise to an expanded Hilbert space $S_T=S\otimes \tilde{S}$, with $S$ being the standard Hilbert space and $\tilde{S}$ the dual (tilde) space. The second basic element of TFD is a Bogoliubov transformation, introducing a rotation in the tilde and non-tilde variables. The main feature of the TFD is that it allows analysis of the time-evolution of the system in addition to the effects of temperature.

This paper is organized as follows. In section II, the electromagnetic field in Very Special Relativity is presented. The field equation for the photon is obtained. This leads to a massive photon. Some attention to the energy-momentum tensor is given. In section III, the TFD formalism is introduced. In section IV, the vacuum expectation value of the energy-momentum tensor in the TFD formalism is discussed. The Stefan-Boltzmann law and the Casimir effect at zero and finite temperature for the electromagnetic field in the VSR are calculated. These results are for a massive photon. However, the standard result is recovered at the limit where the photon mass disappears. In section V, some concluding remarks are presented.

\section{Electromagnetic field in Very Special Relativity}

In the VSR the field strength is defined in terms of the wiggled derivative, i.e.
\bea
\tilde{F}_{\mu\nu}=\tilde{\partial}_\mu A_\nu-\tilde{\partial}_\nu A_\mu,\label{eq1}
\eea
where
\bea
\tilde{\partial}_\mu=\partial_\mu+\frac{m^2}{2}\frac{n_\mu}{n\cdot \partial}\label{eq2}
\eea
is the wiggle derivative with $n_\mu$ being a light-like four-vector that represents the preferred null direction given as $n_\mu=(1,0,0,1)$ and the $m$ parameter sets the scale for the VSR effects. Using eq. (\ref{eq1}) in eq. (\ref{eq2}) the VSR field strength becomes
\bea
\tilde{F}_{\mu\nu}=F_{\mu\nu}+\frac{1}{2}m^2\left[n_\nu\frac{1}{(n\cdot \partial)^2}(n^\alpha F_{\mu\alpha})-n_\mu\frac{1}{(n\cdot \partial)^2}(n^\alpha F_{\nu\alpha})\right].\label{F}
\eea
It is important to note that, $\tilde{F}_{\mu\nu}$ is not Lorentz but SIM(2) invariant. From its definition, the VSR field strength is invariant under the usual gauge transformation
\bea
A_{\alpha}\rightarrow A_\alpha+\partial_\alpha \Lambda.
\eea

Then the VSR gauge Lagrangian is 
\bea
{\cal L}=-\frac{1}{4}\tilde{F}_{\mu\nu}\tilde{F}^{\mu\nu}.
\eea
Using eq. (\ref{F}) this Lagrangian is
\bea
{\cal L}=-\frac{1}{4}F_{\mu\nu}F^{\mu\nu}-\frac{1}{2}m^2(n^\alpha F_{\mu\alpha})\frac{1}{(n\cdot \partial)^2}(n_\rho F^{\mu\rho}).\label{Lag}
\eea
From the Euler-Lagrange equation leads to
\bea
\partial_\mu\frac{\partial {\cal L}}{\partial(\partial_\mu A_\nu)}-\frac{\partial{\cal L}}{\partial A_\nu}=0
\eea
the equation of motion for $F_{\mu\nu}$ is 
\bea
\partial_\mu F^{\mu\nu}+m^2n^\nu\frac{1}{(n\cdot\partial)^2}\partial_\mu(n_\rho F^{\mu\rho})+m^2\frac{1}{(n\cdot\partial)}(n_\rho F^{\rho\nu})=0.
\eea
In terms of the gauge field, the equation of motion becomes
\bea
\partial^2 A^\nu-\partial^\nu\partial_\mu A^\mu&+&m^2n^\nu\frac{1}{(n\cdot\partial)^2}\left[\partial^2(n\cdot A)-(n\cdot \partial)\partial_\mu A^\mu\right]\nonumber\\
&+&m^2\frac{1}{(n\cdot\partial)}\left[(n\cdot \partial)A^\nu-\partial^\nu(n\cdot A)\right]=0,
\eea
where $\partial^2=\partial_\mu\partial^\mu$. Using the Lorentz gauge $\partial_\mu A^\mu=0$ and imposing the additional condition $n\cdot A=0$, the field equation is 
\bea
\left(\partial^2+m^2\right)A^\nu=0.
\eea
This equation implies that $A^\nu$ is a field with mass $m$. Therefore, in VSR the photon mass coming from a term that is gauge invariant, unlike the case where a term of the type $m^2 A^\mu A_\mu$ is not gauge invariant. In addition, the VSR gauge symmetry provides a suitable framework to describe massive modes 
as in the standard case, i.e., without changing the number of physical polarization states of the photon \cite{Cheon, Rivelles}.

Another important ingredient for the study developed here is the energy-momentum tensor associated with the electromagnetic field
\bea 
\mathbb{T}^{\mu\nu}=\frac{\partial {\cal L}}{\partial(\partial_\mu A_\lambda)}\partial^\nu A_\lambda-\eta^{\mu\nu} {\cal L}.
\eea
Then using the Lagrangian (\ref{Lag}) leads to
\bea
\mathbb{T}^{\mu\nu}&=&-F^{\mu\lambda}\partial^\nu A_\lambda+ \frac{1}{4}\eta^{\mu\nu}F_{\rho\sigma}F^{\rho\sigma}-m^2\Bigl\{\Bigl[n^\lambda\frac{1}{(n\cdot \partial)^2}(n_\rho F^{\mu\rho})-n^\mu\frac{1}{(n\cdot \partial)^2}(n_\rho F^{\lambda\rho})\Bigl]\partial^\nu A_\lambda\nonumber\\
&-&\frac{\eta^{\mu\nu}}{2}(n^\alpha F_{\lambda\alpha})\frac{1}{(n\cdot \partial)^2}(n_\rho F^{\lambda\rho})\Bigl\}.
\eea
This tensor is not symmetric. In order to obtain a gauge invariant and symmetric energy-momentum tensor, let's to make $F^{\mu\lambda}\partial^\nu A_\lambda=\eta^{\mu\rho}F_{\rho\lambda}\partial^\nu A^\lambda$ and $F^{\lambda\rho}\partial^\nu A_\lambda=\eta^{\rho\delta}F_{\lambda\delta}\partial^\nu A^\lambda$. Then
\bea
\mathbb{T}^{\mu\nu}&=&-\eta^{\mu\rho}F_{\rho\lambda}F^{\nu\lambda}-\eta^{\mu\rho}F_{\rho\lambda}\partial^\lambda A^\nu+\frac{1}{4}\eta^{\mu\nu}F_{\rho\sigma}F^{\rho\sigma}-m^2\Bigl\{\Bigl[n_\lambda\frac{1}{(n\cdot \partial)^2}(n_\rho F^{\mu\rho})\nonumber\\
&-&\eta^{\rho\delta}n^\mu\frac{1}{(n\cdot \partial)^2}(n_\rho F_{\lambda\delta})\Bigl]F^{\nu\lambda}-\frac{1}{2}\eta^{\mu\nu}(n^\alpha F_{\lambda\alpha})\frac{1}{(n\cdot \partial)^2}(n_\rho F^{\lambda\rho})\Bigl\}\nonumber\\
&-&m^2\Bigl[n_\lambda\frac{1}{(n\cdot \partial)^2}(n_\rho F^{\mu\rho})-\eta^{\rho\delta}n^\mu\frac{1}{(n\cdot \partial)^2}(n_\rho F_{\lambda\delta})\Bigl]\partial^\lambda A^\nu.
\eea
Using the Belinfante method \cite{Belinfante}, a new tensor is defined as
\bea
T^{\mu\nu}=\mathbb{T}^{\mu\nu}+\eta^{\mu\rho}F_{\rho\lambda}\partial^\lambda A^\nu+m^2\Bigl[n_\lambda\frac{1}{(n\cdot \partial)^2}(n_\rho F^{\mu\rho})-\eta^{\rho\delta}n^\mu\frac{1}{(n\cdot \partial)^2}(n_\rho F_{\lambda\delta})\Bigl]\partial^\lambda A^\nu.
\eea
Then the energy-momentum tensor for the electromagnetic field in VSR is 
\bea
T^{\mu\nu}&=&-F^{\mu}\,_{\lambda}F^{\nu\lambda}+\frac{1}{4}\eta^{\mu\nu}F_{\rho\sigma}F^{\rho\sigma}-m^2\Bigl[n_\lambda\frac{1}{(n\cdot \partial)^2}(n_\rho F^{\mu\rho})F^{\nu\lambda}\nonumber\\
&-&\eta^{\rho\delta}n^\mu\frac{1}{(n\cdot \partial)^2}(n_\rho F_{\lambda\delta})F^{\nu\lambda}]-\frac{1}{2}\eta^{\mu\nu}(n^\alpha F_{\lambda\alpha})\frac{1}{(n\cdot \partial)^2}(n_\rho F^{\lambda\rho})\Bigl].
\eea
It is important to note that this tensor is not completely symmetric. This is a feature of theories which exhibit Lorentz violation. However, at the limit $m^2=0$ the standard electromagnetic energy-momentum tensor is recovered.

In order to avoid divergences, the energy-momentum tensor is written at different space-time points as
\bea
T^{\mu\nu}(x)&=&\lim_{x'\rightarrow x}\Bigl[-{\cal F}^{\mu\lambda,\nu}\,_\lambda(x,x')+\frac{1}{4}\eta^{\mu\nu}{\cal F}_{\rho\sigma}\,^{\rho\sigma}(x,x')-\frac{m^2}{(n\cdot\partial)^2}\Bigl(n_\lambda n_\rho{\cal F}^{\mu\rho,\nu\lambda}(x,x')\nonumber\\
&-& n^\mu n^\delta {\cal F}_{\lambda\delta,}\,^{\nu\lambda}(x,x')-\frac{1}{2}\eta^{\mu\nu}n^\delta n_\rho {\cal F} _{\lambda\delta,}\,^{\lambda\rho}(x,x')\Bigl)\Bigl]
\eea
with
\bea
{\cal F}^{\mu\nu,\lambda\rho}(x,x')&=&\tau\left[F^{\mu\nu}(x)F^{\lambda\rho}(x')\right]\nonumber\\
&=&F^{\mu\nu}(x)F^{\lambda\rho}(x')\theta(x_0-x_0')+F^{\lambda\rho}(x')F^{\mu\nu}(x)\theta(x_0'-x_0),\label{F2}
\eea
where $\tau$ is the time-order operator and $\theta(x_0-x_0')$ is the step function. In calculations that follow, let us consider the relation
\bea
\partial^\mu\theta(x_0-x_0')=v_0^\mu\delta(x_0-x_0')
\eea
with $v_0^\mu$ being the $\mu$-component of the time-like vector $v_0=(1,0,0,0)$, and the commutation relation
\bea
[A_i(x), \pi_j(x')]=i\left(\delta_{ij}-\frac{\partial_i\partial_j}{\nabla^2}\right)\delta(x-x'),
\eea
where $\pi_j(x)$ is the canonical conjugate momentum. Then eq. (\ref{F2}) becomes
\bea
{\cal F}^{\mu\nu,\lambda\rho}(x,x')&=&\Gamma^{\mu\nu,\lambda\rho,\alpha\gamma}(x,x')\tau[A_\alpha(x)A_\gamma(x')]+I^{\nu,\lambda\rho}(x,x')v_0^\mu\delta(x_0-x_0')\nonumber\\
&-&I^{\mu,\lambda\rho}(x,x')v_0^\nu\delta(x_0-x_0').
\eea
Here the first term is defined as
\bea
\Gamma^{\mu\nu,\lambda\rho,\alpha\gamma}(x,x')\equiv \left(\eta^{\nu\alpha}\partial^\mu-\eta^{\mu\alpha}\partial^\nu\right)\left(\eta^{\rho\gamma}\partial'^\lambda-\eta^{\lambda\gamma}\partial'^\rho\right)
\eea
and
\bea
I^{\mu,\lambda\nu}(x,x')\equiv iv_0^\lambda\left(\eta^{\mu\nu}-\nabla^{-2}\partial^\mu\partial^\nu\right)\delta(x-x')-iv_o^\nu\left(\eta^{\mu\lambda}-\nabla^{-2}\partial^\mu\partial^\lambda\right)\delta(x-x').
\eea
Using these definitions, the energy-momentum tensor becomes
\bea
T^{\mu\nu}(x)&=&\lim_{x'\rightarrow x}\Bigl\{-\Delta^{\mu\nu,\alpha\gamma}(x,x')\tau[A_\alpha(x)A_\gamma(x')]+\Pi^{\mu\nu}\delta(x-x')\nonumber\\
&-&\frac{m^2}{(n\cdot\partial)^2}\left(\Delta_m^{\mu\nu,\alpha\gamma}(x,x')\tau[A_\alpha(x)A_\gamma(x')]+\Pi_m^{\mu\nu}\delta(x-x')\right)\Bigl\},
\eea
where
\bea
\Delta^{\mu\nu,\alpha\gamma}(x,x')&=&\Gamma^{\mu\lambda,\nu}\,_{\lambda,}\,^{\alpha\gamma}(x,x')-\frac{1}{4}\eta^{\mu\nu}\Gamma^{\rho\sigma,}\,_{\rho\sigma,}\,^{\alpha\gamma}(x,x'),\\
\Pi^{\mu\nu}&=&2iv_0^\mu v_0^\nu-\frac{3}{2}i\eta^{\mu\nu},\\
\Delta_m^{\mu\nu,\alpha\gamma}(x,x')&=&n_\lambda n_\rho\Gamma^{\mu\rho, \nu\lambda,\alpha\gamma}(x,x')-n^\mu n^\delta\Gamma_{\lambda\delta,}\,^{\nu\lambda,}\,^{\alpha\gamma}(x,x')-\frac{1}{2}\eta^{\mu\nu}n^\delta n_\rho\Gamma_{\lambda\delta,}\,^{\lambda\rho,\alpha\gamma}(x,x'),\\
\Pi_m^{\mu\nu}&=&in_\lambda n_\rho\left(v_0^\nu v_0^\mu\eta^{\rho\lambda}-v_0^\lambda v_0^\mu\eta^{\rho\nu}-v_0^\nu v_0^\rho\eta^{\mu\lambda}+v_0^\lambda v_0^\rho\eta^{\mu\nu}\right)\nonumber\\
&-&in^\mu n^\delta(2v_0^\nu v_{0\delta}-\eta_\delta^\nu)-\frac{1}{2}\eta^{\mu\nu}n^\delta n_\rho i(\eta_\delta^\rho-2v_0^\rho v_{0\delta}).
\eea

Now let us calculate the vacuum expectation value of $T^{\mu\nu}(x)$. It is an important quantity that will be used in some applications in the next sections. Then
\bea
\langle T^{\mu\nu}(x)\rangle &=&\langle 0|T^{\mu\nu}(x)|0\rangle\nonumber\\
&=&\lim_{x'\rightarrow x}\Bigl\{-\Delta^{\mu\nu,\alpha\gamma}(x,x')\langle 0|\tau[A_\alpha(x)A_\gamma(x')]|0\rangle+\Pi^{\mu\nu}\delta(x-x')\nonumber\\
&-&\frac{m^2}{(n\cdot\partial)^2}\left(\Delta_m^{\mu\nu,\alpha\gamma}(x,x')\langle 0|\tau[A_\alpha(x)A_\gamma(x')]|0\rangle+\Pi_m^{\mu\nu}\delta(x-x')\right)\Bigl\}.
\eea
It is known that
\bea
\langle 0|\tau[A_\alpha(x)A_\gamma(x')]|0\rangle=D_{\alpha\gamma}(x-x'),
\eea
with $D_{\alpha\gamma}(x-x')$ being the propagator for the gauge field in VSR. It is to be noted that the photon is massive. Then in the momentum space 
\bea
D_{\alpha\gamma}(p)=-i\frac{\eta_{\alpha\gamma}}{p^2-m^2}.
\eea
This propagator is written as
\bea
D_{\alpha\gamma}(x-x')=-i\eta_{\alpha\gamma}G_0(x-x'),
\eea
where $G_0(x-x')$ is the massive scalar field propagator 
\bea
G_0(x-x')=-\frac{i m}{4\pi^2}\frac{K_1(m\sqrt{-(x-x')^2})}{\sqrt{-(x-x')^2}}
\eea
with $K_\nu(Z)$ being the modified Bessel function. 

Then the vacuum expectation value of the energy-momentum tensor becomes
\bea
\langle T^{\mu\nu}(x)\rangle &=&\lim_{x'\rightarrow x}\Bigl[i \Gamma^{\mu\nu}G_0(x-x')+\Pi^{\mu\nu}\delta(x-x')\nonumber\\
&+&\frac{m^2}{(n\cdot\partial)^2}\Bigl(i \Gamma_m^{\mu\nu}G_0(x-x')+\Pi_m^{\mu\nu}\delta(x-x')\Bigl)\Bigl],\label{VEV}
\eea
where
\bea
\Gamma^{\mu\nu}=2\left(\partial^\mu\partial'^\nu-\frac{1}{4}\eta^{\mu\nu}\partial^\rho\partial'_\rho\right)
\eea
and 
\bea
\Gamma_m^{\mu\nu}=-n_\lambda n^\nu\partial^\mu\partial'^\lambda+n^\mu n_\rho\partial^\rho\partial'^\nu+n^\mu n^\nu\partial_\lambda\partial'^\lambda.
\eea

In the next section, the TFD formalism is introduced and then some applications are investigated from the vacuum expectation value of the energy-momentum tensor associated with the electromagnetic field.

\section{Thermo Field Dynamics formalism}

An introduction to the Thermo Field Dynamics (TFD) formalism, which is a real-time finite temperature field theory, is presented. This based on two main ingredients: (i) the Hilbert space ${\cal S}$ is doubled leading to an expanded space which is defined as ${\cal S}_T={\cal S}\otimes \tilde{\cal S}$, with $\tilde{\cal S}$ being the dual (tilde) Hilbert space. The duplicate space is constructed using the tilde ($^\thicksim$) conjugation rules, associating each operator in ${\cal S}$ to two operators in ${\cal S}_T$. (ii) The Bogoliubov transformation is the other key ingredient required by the TFD. Such transformation introduces thermal effects through a rotation between tilde ($\tilde{\cal S}$) and non-tilde (${\cal S}$) operators. The Bogoliubov transformation, ${\cal U}(\alpha)$, is defined as
\bea
{\cal U}(\alpha)=\left( \begin{array}{cc} u(\alpha) & -w(\alpha) \\
\xi w(\alpha) & u(\alpha) \end{array} \right),
\eea
with $u^2(\alpha)+\xi w^2(\alpha)=1$ and the $\alpha$ parameter is assumed as the compactification parameter defined by $\alpha=(\alpha_0,\alpha_1,\cdots\alpha_{D-1})$. The effect of temperature is described by the choice $\alpha_0\equiv\beta$ and $\alpha_1,\cdots\alpha_{D-1}=0$, where $\beta=1/k_B T$ with $k_B$ being the Boltzmann constant.

In order to apply the Bogoliubov transformation, arbitrary operators ${\cal O}$ and $\tilde{\cal O}$ in Hilbert space ${\cal S}$ and $\tilde{\cal S}$ respectively, are considered. Then
\bea
\left( \begin{array}{cc} {\cal O}(\alpha,k)  \\\xi \tilde {\cal O}^\dagger(\alpha,k) \end{array} \right)={\cal U}(\alpha)\left( \begin{array}{cc} {\cal O}(k)  \\ \xi\tilde {\cal O}^\dagger(k) \end{array} \right),
\eea
where $\xi = -1$ for bosons and $\xi = +1$ for fermions. 

It is important to note that, in this work the topology $\Gamma_D^d=(\mathbb{S}^1)^d\times \mathbb{R}^{D-d}$ with $1\leq d \leq D$, is considered. Here $D$ are the space-time dimensions and $d$ is the number of compactified dimensions. In addition, in this formalism any set of dimensions of the manifold $\mathbb{R}^{D}$ can be compactified.

It is observed that any propagator in the TFD formalism can be written in terms of the $\alpha $-parameter. For the scalar field we get \cite{Umezawa2}
\bea
G_0^{(AB)}(x-x';\alpha)=i\langle 0,\tilde{0}| \tau[\phi^A(x;\alpha)\phi^B(x';\alpha)]| 0,\tilde{0}\rangle,
\eea
where,
\bea
\phi(x;\alpha)&=&{\cal U}(\alpha)\phi(x){\cal U}^{-1}(\alpha),
\eea
with $A$ and $B$ $=1-2$. Considering the thermal vacuum $|0(\alpha)\rangle={\cal U}(\alpha)|0,\tilde{0}\rangle$, the propagator becomes
\bea
G_0^{(AB)}(x-x';\alpha)&=&i\langle 0(\alpha)| \tau[\phi^A(x)\phi^B(x')]| 0(\alpha)\rangle,\nonumber\\
&=&i\int \frac{d^4k}{(2\pi)^4}e^{-ik(x-x')}G_0^{(AB)}(k;\alpha),
\eea
where
\bea
G_0^{(AB)}(k;\alpha)={\cal U}^{-1}(\alpha)G_0^{(AB)}(k){\cal U}(\alpha),
\eea
with
\bea
G_0^{(AB)}(k)=\left( \begin{array}{cc} G_0(k) & 0 \\
0 & \xi G^*_0(k) \end{array} \right),
\eea
and
\bea
G_0(k)=\frac{1}{k^2-m^2+i\epsilon}.
\eea
Here $m$ is the scalar field mass.

By taking $A=B=1$, since the physical quantities are given by the non-tilde variables, the Green function is given as
\bea
G_0^{(11)}(k;\alpha)=G_0(k)+\xi w^2(k;\alpha)[G^*_0(k)-G_0(k)],
\eea
where $w^2(k;\alpha)$ is the generalized Bogoliubov transformation \cite{GBT} which is given as
\bea
w^2(k;\alpha)=\sum_{s=1}^d\sum_{\lbrace\sigma_s\rbrace}2^{s-1}\sum_{l_{\sigma_1},...,l_{\sigma_s}=1}^\infty(-\xi)^{s+\sum_{r=1}^sl_{\sigma_r}}\,\exp\left[{-\sum_{j=1}^s\alpha_{\sigma_j} l_{\sigma_j} k^{\sigma_j}}\right],\label{BT}
\eea
with $d$ being the number of compactified dimensions, $\lbrace\sigma_s\rbrace$ denotes the set of all combinations with $s$ elements and $k$ is the 4-momentum. 

\section{Stefan-Boltzmann law and Casimir effect}

In this section, some applications are investigated. To achieve this goal, the vacuum expectation value of the energy-momentum tensor, eq. (\ref{VEV}), is written in the context of TFD. Following the tilde conjugation rules, the vacuum average of the energy-momentum tensor reads
\bea
\langle T^{\mu\nu(AB)}(x,\alpha)\rangle &=&\lim_{x'\rightarrow x}\Bigl[i \Gamma^{\mu\nu}G_0^{(AB)}(x-x';\alpha)+\Pi^{\mu\nu}\delta(x-x')\nonumber\\
&+&\frac{m^2}{(n\cdot\partial)^2}\Bigl(i \Gamma_m^{\mu\nu}G_0^{(AB)}(x-x';\alpha)+\Pi_m^{\mu\nu}\delta(x-x')\Bigl)\Bigl].\label{VEV1}
\eea
In order to estimate the effects of the topology and inspired by the usual Casimir prescription, a measurable physical quantity is given by
\bea
{\cal T}^{\mu\nu(AB)}(x,\alpha)=i\lim_{x'\rightarrow x}\Bigl[ \Gamma^{\mu\nu}\overline{G}_0^{(AB)}(x-x';\alpha)+\frac{m^2}{(n\cdot\partial)^2}\Gamma_m^{\mu\nu}\overline{G}_0^{(AB)}(x-x';\alpha)\Bigl],\label{Final}
\eea
where
\bea
{\cal T}^{\mu\nu(AB)}(x;\alpha)\equiv\left\langle T^{\mu\nu(AB)}(x;\alpha)\right\rangle-\left\langle T^{\mu\nu(AB)}(x)\right\rangle,
\eea
and
\bea
\overline{G}_0^{(AB)}(x-x';\alpha)=G_0^{(AB)}(x-x';\alpha)-G_0^{(AB)}(x-x').\label{Green}
\eea

Now the energy-momentum tensor given in eq. (\ref{Final}) will be analyzed for three different topologies. (i) The topology $\Gamma_4^1=\mathbb{S}^1\times\mathbb{R}^{3}$, where $\alpha=(\beta,0,0,0,)$. In this case the time-axis is compactified in $\mathbb{S}^1$, with circumference $\beta$. (ii) The topology $\Gamma_4^1$ with $\alpha=(0,0,0,i2d)$, where the compactification along the coordinate $z$ is considered. (iii) The topology $\Gamma_4^2=\mathbb{S}^1\times\mathbb{S}^1\times\mathbb{R}^{2}$ with $\alpha=(\beta,0,0,i2d)$ is used. In this case, there is a double compactification that consists of one being the time and the other along the coordinate $z$.

\subsection{Stefan-Boltzmann law}

Here is used $\alpha=(\beta,0,0,0,)$. Then the generalized Bogoliubov transformation, eq. (\ref{BT}), becomes
\bea
w^2(\beta)=\sum_{l_0=1}^{\infty}e^{-\beta k^0l_0}\label{BT1}
\eea
and the Green function is given as
\bea
\overline{G}_0(x-x';\beta)=2\sum_{l_0=1}^{\infty}G_0(x-x'-i\beta l_0n_0),\label{GF1}
\eea
where $n_0=(1,0,0,0)$. The energy-momentum tensor reads
\bea
{\cal T}^{\mu\nu(AB)}(x,\beta)=2i\lim_{x'\rightarrow x}\sum_{l_0=1}^{\infty}\Bigl[ \Gamma^{\mu\nu}+\frac{m^2}{(n\cdot\partial)^2}\Gamma_m^{\mu\nu}\Bigl]G_0(x-x'-i\beta l_0n_0).
\eea
By taking $\mu=\nu=0$ we get
\bea
{\cal T}^{00(11)}(\beta)=\frac{m}{4\pi^2}\sum_{l_0=1}^{\infty}\left[\frac{12m}{(\beta l_0)^2}K_0(m\beta l_0)+\frac{\left(24(n\cdot\partial)^2+(ml_0)^2(3(n\cdot\partial)^2+2m^2)\right)\beta^2}{(n\cdot\partial)^2(\beta l_0)^3}K_1(m\beta l_0)\right].
\eea
This is the Stefan-Boltzmann-type law in VSR. It is interesting to note that, at the limit $m\rightarrow 0$ the standard Stefan-Boltzmann law is recovered, i.e.,
\bea
E_{SB}=\frac{\pi^2}{15}T^4,
\eea
where $E_{SB}={\cal T}^{00(11)}_{m\rightarrow 0}(\beta)$. Therefore, the usual Stefan-Boltzmann law, i.e. $E\sim T^4$, is changed due to the mass of the photon.

\subsection{Casimir effect at zero temperature}

In this case, $\alpha=(0,0,0,i2d)$ is considered. Then
\bea
w^2(d)=\sum_{l_3=1}^{\infty}e^{-i2d k^3l_3}\label{BT2}
\eea
is the Bogoliubov transformation and 
\bea
\overline{G}_0(x-x';d)=2\sum_{l_3=1}^{\infty}G_0(x-x'-2d l_3n_3)\label{GF2}
\eea
is the Green function with $n_3=(0,0,0,1)$. Thus the energy-momentum tensor is given as
\bea
{\cal T}^{\mu\nu(AB)}(x,d)=2i\lim_{x'\rightarrow x}\sum_{l_3=1}^{\infty}\Bigl[ \Gamma^{\mu\nu}+\frac{m^2}{(n\cdot\partial)^2}\Gamma_m^{\mu\nu}\Bigl]G_0(x-x'-2d l_3n_3).
\eea

For $\mu=\nu=0$, the Casimir energy in VSR is found as
\bea
{\cal T}^{00(11)}(d)=-\frac{m}{8\pi^2}\sum_{l_3=1}^{\infty}\left[\frac{2m}{(d l_3)^2}K_0(2md l_3)+\frac{\left(-2(dm^2l_3)^2+(n\cdot\partial)^2(2+(dml_3)^2)\right)}{(n\cdot\partial)^2(d l_3)^3}K_1(2dm l_3)\right].
\eea
And for $\mu=\nu=3$ the Casimir pressure in VSR is
\bea
{\cal T}^{33(11)}(d)=-\frac{m}{8\pi^2}\sum_{l_3=1}^{\infty}\left[\frac{6m}{(d l_3)^2}K_0(2md l_3)+\frac{\left(-2(dm^2l_3)^2+3(n\cdot\partial)^2(2+(dml_3)^2)\right)}{(n\cdot\partial)^2(d l_3)^3}K_1(2dm l_3)\right].
\eea

By taking the limit $m\rightarrow 0$ and defining $E_c={\cal T}^{00(11)}_{m\rightarrow 0}(d)$ and $P_c={\cal T}^{33(11)}_{m\rightarrow 0}(d)$, are obtained
\bea
E_c&=&-\frac{\pi^2}{720d^4},\\
P_c&=&-\frac{\pi^2}{240d^4}.
\eea
These are the usual Casimir energy and pressure for the electromagnetic field. 

\subsection{Casimir effect at finite temperature}

Now $\alpha=(\beta,0,0,i2d)$ is considered. Then the Casimir effect at finite temperature and with spatial compactification is calculated. In this case, the Bogoliubov transformation is
\bea
w^2(\beta,d)=\sum_{l_0=1}^\infty(-1)^{1+l_0}e^{-\beta k^0l_0}+\sum_{l_3=1}^\infty(-1)^{1+l_3}e^{-i2d k^3l_3}+2\sum_{l_0,l_3=1}^\infty(-1)^{l_0+l_3}e^{-\beta k^0l_0-i2d k^3l_3}.
\eea
Here the first two terms correspond to the Stefan-Boltzmann law and the Casimir effect at zero temperature. The Casimir energy and Casimir pressure at finite temperature are calculated using the third term. For the third term the Green function is
\bea
\overline{G}_0(x-x';\beta, d)=4\sum_{l_0,l_3=1}^{\infty}G_0(x-x'-i\beta l_0n_0-2d l_3n_3).\label{GF3}
\eea

Then the energy-momentum tensor becomes
\bea
{\cal T}^{\mu\nu(AB)}(x;\beta,d)=4i\lim_{x'\rightarrow x}\sum_{l_0,l_3=1}^{\infty}\Bigl[ \Gamma^{\mu\nu}+\frac{m^2}{(n\cdot\partial)^2}\Gamma_m^{\mu\nu}\Bigl]G_0(x-x'-i\beta l_0n_0-2d l_3n_3).
\eea
Thus the Casimir energy at finite temperature is given as
\bea
{\cal T}^{00(11)}(\beta,d)&=&-\sum_{l_0,l_3=1}^{\infty}\frac{m}{2\pi^2(n\cdot\partial)^2[(2dl_3)^2+(\beta l_0)^2]^{5/2}}\Bigl\{4m(n\cdot\partial)^2[(2dl_3)^2+(\beta l_0)^2]^{1/2}\\
&\times&\bigl((2dl_3)^2-3(\beta l_0)^2\bigl)K_0(m\sqrt{(2dl_3)^2+(\beta l_0)^2})+\bigl[2m^4\bigl((2dl_3)^2+(\beta l_0)^2\bigl)^2\nonumber\\
&+&(n\cdot\partial)^2\bigl((2dl_3)^2-3(\beta l_0)^2\bigl)\bigl(8+m^2\bigl((2dl_3)^2+(\beta l_0)^2)\bigl)\bigl)\bigl]K_1(m\sqrt{(2dl_3)^2+(\beta l_0)^2})\Bigl\}\nonumber
\eea
and Casimir pressure at finite temperature is
\bea
{\cal T}^{33(11)}(\beta,d)&=&-\sum_{l_0,l_3=1}^{\infty}\frac{m}{2\pi^2(n\cdot\partial)^2[(2dl_3)^2+(\beta l_0)^2]^{5/2}}\Bigl\{4m(n\cdot\partial)^2[(2dl_3)^2+(\beta l_0)^2]^{1/2}\\
&\times&\bigl(3(2dl_3)^2-(\beta l_0)^2\bigl)K_0(m\sqrt{(2dl_3)^2+(\beta l_0)^2})+\bigl[2m^4\bigl((2dl_3)^2+(\beta l_0)^2\bigl)^2\nonumber\\
&+&(n\cdot\partial)^2\bigl(3(2dl_3)^2-(\beta l_0)^2\bigl)\bigl(8+m^2\bigl((2dl_3)^2+(\beta l_0)^2)\bigl)\bigl)\bigl]K_1(m\sqrt{(2dl_3)^2+(\beta l_0)^2})\Bigl\}\nonumber.
\eea
At the limit $m\rightarrow 0$, the standard limit of QED is recovered, that is,
\bea
{\cal T}^{00(11)}_{m\rightarrow 0}(x;\beta,d)=\frac{4}{\pi^2}\sum_{l_0,l_3=1}^{\infty}\frac{3(\beta l_0)^2-(2dl_3)^2}{[(\beta l_0)^2+(2dl_3)^2]^3}
\eea
and
\bea
{\cal T}^{33(11)}_{m\rightarrow 0}(x;\beta,d)=\frac{4}{\pi^2}\sum_{l_0,l_3=1}^{\infty}\frac{(\beta l_0)^2-3(2dl_3)^2}{[(\beta l_0)^2+(2dl_3)^2]^3}.
\eea

Therefore, our results show that the VSR gives mass for the photon. This leads to changes in important results of the usual QED, such as Stefan-Boltzmann law and Casimir effect at zero and finite temperature.

\section{Conclusion} 

The Lorentz symmetry is in good agreement with the experiments, i.e. no violation of this fundamental symmetry has been observed. However, models that seek a fundamental theory that unifies SM and general relativity predict a break in Lorentz invariance at Planck scale energy. Among the enormous class of models that investigate the Lorentz symmetry violation, here the VSR is considered. The main characteristic of the VSR proposal is that the laws of physics are not invariant under the whole Poincaré group but rather under subgroups of the Poincaré group preserving the basic elements of special relativity. In this work, the field equation for the electromagnetic field in VSR is revised. It is shown that in VSR the photon mass coming from a term that is gauge invariant. The energy-momentum tensor for this field is calculated. The non-local term in the energy-momentum tensor shows that this quantity is not symmetric. It is a characteristic of all models that exhibit Lorentz violation. Then the TFD formalism, a real-time finite temperature quantum field theory, is introduced. The Stefan-Boltzmann law and the Casimir effect at zero and finite temperature are calculated for a massive photon. Our results show that the mass of the photon completely change these QED results. However, at the appropriate limit, the standard results for massless photons are recovered. It is important to note that, our results fill a gap in the vast list of effects due to VSR. Therefore, these results, the Stefan-Boltzmann-type law and the Casimir effect for a massive photon are completely new results in the literature. In addition, these results indicate that the combined effect of temperature and spatial compactification give a new constraint for the Casimir effect in VSR.

\section*{Acknowledgments}

This work by A. F. S. is supported by CNPq projects 308611/2017-9 and 430194/2018-8.

\end{document}